\documentclass{article}
 \pdfoutput=1
\usepackage{authblk}
\usepackage[utf8]{inputenc}
\usepackage{diagbox}
\usepackage{amsmath}
\usepackage{amsfonts}
\usepackage{graphicx}
\usepackage{mathtools}
\usepackage{float} 
\usepackage{xcolor}
\usepackage{hyperref}
\usepackage{verbatim}
\usepackage{float}
\providecommand{\keywords}[1]
{
  \small	
  \textbf{\textit{Keywords---}} #1
}

\title{Inverse resolution of spatially varying diffusion coefficient using Physics-Informed neural networks}
\date{}
\author[1]{Sukirt Thakur}

\affil[1]{\small{School of Mechanical Engineering,
            Purdue University, 
            West Lafayette,
            47907, 
            Indiana,
            USA}}
\author[1]{Ehsan  Esmaili}

\author[2]{Sarah Libring}

\affil[2]{\small{ Weldon School of Biomedical Engineering,
            Purdue University, 
            West Lafayette,
            47907, 
            Indiana,
            USA}}
\author[2]{Luis Solorio}
\author[1]{Arezoo M. Ardekani}

\begin{document}
\maketitle 

\begin{abstract} 
{Resolving the diffusion coefficient is a key element in many biological and engineering systems, including pharmacological drug transport and fluid mechanics analyses. Additionally, these systems often have spatial variation in the diffusion coefficient which must be determined, such as for injectable drug-eluting implants into heterogeneous tissues. Unfortunately, obtaining the diffusion coefficient from images in such cases is an inverse problem with only discrete data points. The development of a robust method that can work with such noisy and ill-posed datasets to accurately determine spatially-varying diffusion coefficients is of great value across a large range of disciplines. Here, we developed an inverse solver that uses physics informed neural networks (PINNs) to calculate spatially-varying diffusion coefficients from numerical and experimental image data in varying biological and engineering applications. The residual of  the transient diffusion equation for a concentration field is minimized to find the diffusion coefficient. The robustness of the method as an inverse solver was tested using both numerical and experimental datasets. The predictions show good agreement with both the numerical and experimental benchmarks; an error of less than 6.31\% was obtained against all numerical benchmarks, while the diffusion coefficient calculated in experimental datasets matches the appropriate ranges of other reported literature values. Our work demonstrates the potential of using PINNs to resolve spatially-varying diffusion coefficients, which may aid a wide-range of applications, such as enabling better-designed drug-eluting implants for regenerative medicine or oncology fields.}

\end{abstract}

\section{Introduction}
Understanding the spatial variation of the diffusion coefficient is a key element in many biological and engineering systems (e.g., drug transport, biomedical transport, petroleum engineering, and hydrogeology) \cite{bruna2015,ning2017}. The diffusion coefficient is a proportionality constant which relates the gradient in concentration of a species with the mass flux and gives a measure of the rate of diffusion. However, there are numerous hurdles when attempting to resolve the diffusion coefficient from experimental data. First, the diffusion coefficient often varies spatially, sometimes discontinuously, due to heterogeneity in biological tissues, porous media, and soil. Second, determining the diffusion coefficient from experimental images is an inverse problem, where the parameters of the system need to be determined from the given observations. Numerical instability is a challenge for such problems, which are fundamentally ill-posed in the Hadamard sense \cite{Kabanikhin2008}. Even a small perturbation in experiments can significantly change the estimated diffusion coefficient \cite{Yang2008}. Additionally, data is always collected at discrete points, such that any errors in observation will be amplified in the parameters to be determined \cite{Kern2016}. Moreover, multiple models can fit the data and uniqueness of the solution for the inverse problem is not guaranteed. A robust method which can work with noisy data to accurately determine the diffusion coefficient has wide-ranging applications and is of great value across a large range of disciplines. Many numerical methods replace the original inverse problem by linearization around constant coefficients in order to solve a simpler problem \cite{Isakov2006}. For example, numerical methods such as finite differences in combination with least squares \cite{Shifdar2006}, genetic algorithm with sinc-Galerkin method \cite{Mazraeh2017}, as well as Bayesian methods \cite{Fudym2008,boodaghi2021} have been used to estimate the diffusion coefficient for the observed data. However, obtaining numerical solutions that are satisfactory using such methods is often difficult \cite{Isakov2006}. 

Neural networks are powerful computational tools that have the capacity to represent complex linear and non-linear systems \cite{Lusch2018}. There is a growing interest in using neural networks to extract latent information from data in engineering fields like fluid mechanics \cite{Brunton2020} and material engineering \cite{Sanchez2018}. Machine learning techniques have been used to obtain corrective terms for improving closure models in computational physics through inverse modelling of high-fidelity simulation and experimental data \cite{Parish2016,Ling2016}. There is a growing interest in using neural networks as a tool for better understanding biological systems  such as studying membranes in fluorescence microscopy \cite{Writh2021} and analysis of molecular images \cite{AutoTracer2021}. Although traditional neural networks thrive on data, high-fidelity computational and experimental data is scarce and expensive when it comes to many engineering and biological systems. Raissi et al. \cite{Raissi2017, Raissi2017a, Raissi2019} have introduced physics informed neural networks (PINNs), which are neural networks that respect physical laws described by partial differential equations (PDEs) while performing supervised learning tasks. PINNs are extremely data efficient neural networks that can be used to solve PDEs (forward problems) or to discover the coefficients of terms in the PDE given the solution (inverse problems) \cite{Lu2021}. PINNs have offered a novel solution methodology to diverse problems like brittle fracture mechanics \cite{Goswami2019}, learning constitutive relations for a non-stationary bioreactor \cite{Tipireddy2019}, modelling the thermochemical curing process of composite-tool systems during manufacture \cite{niaki2021}, predicting arterial blood pressure from 4D MRI data \cite{kissas2020}, and modelling soft tissues \cite{liu2020}. The methodology has also been used to solve inverse problems like reconstructing the velocity and pressure fields from flow visualizations \cite{Raissi2020} and discovering turbulence models from noisy spatio-temporal data \cite{Raissi2019a}. The ability of PINNs to work with noisy data makes it a promising option to solve inverse problems using experimental observations.
 
Our focus in this work was to employ PINNs to solve the inverse problem of discovering the spatially varying diffusion coefficient from spatio-temporal information of the diffused passive scalar. Our method was subjected to both numerical and experimental data sets. We first used clean numerical data for verification of our method and then worked with three distinct experimental datasets which contained noise. These results demonstrated that we were able to accurately resolve the diffusion coefficient in different liquids and hydrogels. Additionally, the framework presented is robust and the method can be modified to tackle inverse problems of similar nature with minor modifications. In total, we conclude that this method may be an ideal platform to predict the properties of many substrates in biological systems based on spatially-varying diffusivity, with numerous engineering applications.

\section{Problem definition and methodology} \label{Methodology}

\begin{figure}
        \centering
        \includegraphics[width=\linewidth]{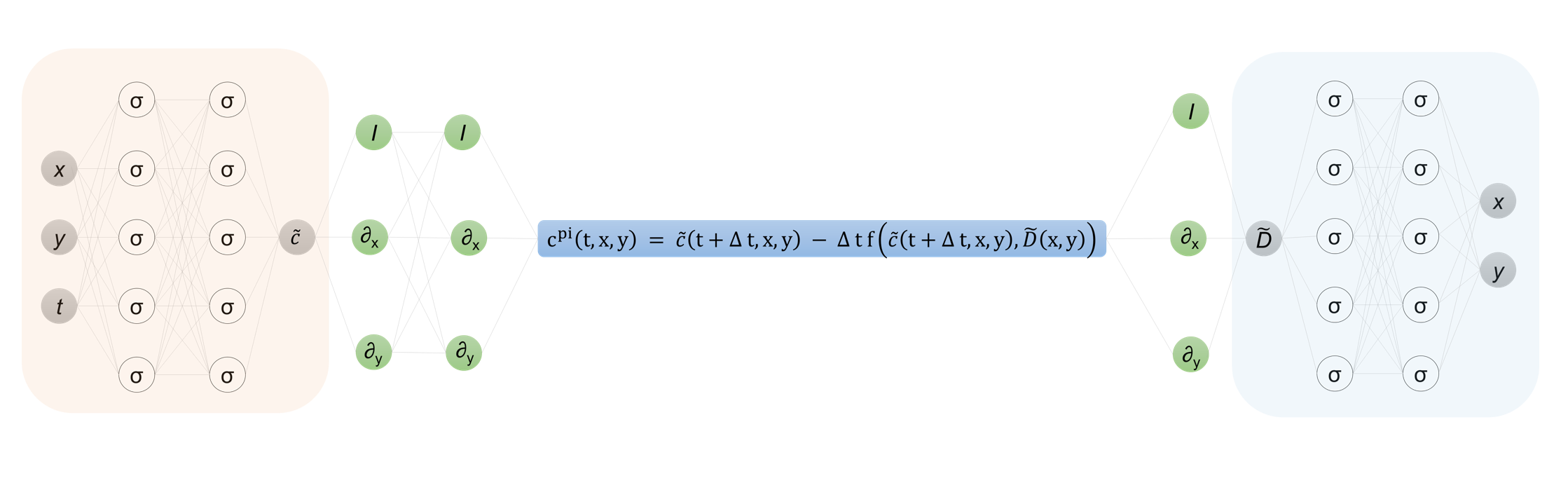}
        \caption{We use two densely connected neural networks which are 8 layers deep for both the concentration ($\Tilde{c}$) and diffusion coefficient($\Tilde{D}$). For the purpose of this schematics, the networks are shown to have 2 hidden layers and 5 neurons per hidden layer. }
        \label{fig:C-Dnetworks}
    \end{figure}

 Our objective was to solve the inverse problem of determining the spatially varying diffusion coefficient which best described a given set of spatio-temporal data of the concentration of a passive scalar using PINNs. To do so, we employed the use of fully connected neural networks. Fully connected neural networks, or feed-forward neural networks, have the simplest architecture where the output of one layer of the neural network is used as the input to the next layer. For a network with an input layer, an output layer, and H hidden layers, the $k^{th}$ hidden layer will receive an output from the previous layer ($x^{k-1}$) and perform the affine transform first
 \begin{equation}
     \mathcal{N}_{k}(x^{k-1})=w^{k}x^{k-1}+b^{k},
 \end{equation}
 where $w^{k}$ and $b^{k}$ are the weights and biases associated with the $k^{th}$ hidden layer. Before sending this output as the input to the next layer, a nonlinear activation function $\sigma$ is applied. For an input $x$, a fully connected neural network can be mathematically written as 
 \begin{equation}
     \mathcal{NN}_{\theta}(x) = (\mathcal{N}_{H} \circ \sigma \circ \mathcal{N}_{H-1} \circ ... \circ \sigma \mathcal{N}_{1})(x),
 \end{equation}
 where $\circ$ is the composition operator and $\theta$ represents the trainable weights and biases of the network which are optimized. The key idea of physics informed networks is to constrain the neural network such that the residual of the pertinent partial differential equation is minimized.
 Consider the transient diffusion equation for a concentration field of a passive scalar $c(x,y,t)$ in two dimensions with a spatially varying diffusion coefficient $D(x,y)$
\begin{equation}\label{DiffusionEqn}
    c_{t} - c_{x}D_{x} - c_{y}D_{y} - D(c_{xx}+c_{yy}) =0.
\end{equation}
This is a parabolic equation, analogous to the equations pertinent to problems in heat transfer and the same framework can be used to determine thermal properties using temperature measurements.  
We approximate the functions $\Tilde{c}(x, y, t)$ and $\Tilde{D} (x, y)$ using two deep fully connected neural networks. The only known data are point clouds inside the domain $\{{x}, {y}, {t}, {c(x,y,t)}\}$. Given this spatio-temporal data, we want to infer the spatially varying diffusion coefficient $D(x,y)$. To solve this inverse problem, we designed a PINN to approximate the diffusion coefficient $\Tilde{D}(x,y)$ which best describes the concentration field $c(x,y,t)$ data. Now, considering \eqref{DiffusionEqn}, we define
\begin{equation}
    f(c, D) = {c}_{x}{D}_{x} - {c}_{y}{D}_{y} - {D}({c}_{xx}+{c}_{yy}),
\end{equation} where the subscripts denote the derivatives. We now create a physics-informed neural network using backward Euler discretization. We define
\begin{equation}\label{residual_eqn}
    c^{pi}(t,x,y;\Delta t, \theta, \phi) = \Tilde{c}(t + \Delta t,x,y; \theta) - \Delta t f(\Tilde{c}(t + \Delta t,x,y; \theta),\Tilde{D}(x,y; \phi)),
\end{equation}
where $\theta$ and $\phi$ are the parameters for the neural networks used to approximate $\Tilde{c}(x, y, t)$ and $\Tilde{D} (x, y)$, respectively. As shown in Fig. \ref{fig:C-Dnetworks}, to create our PINN we obtained $c^{pi}$ corresponding to equation \eqref{DiffusionEqn} using the two neural networks defined above. The identity operator is represented using $I$ and the differential operators are represented using $\partial_{x}$, $\partial_{y}$ and $\partial_{t}$ in Fig. \ref{fig:C-Dnetworks}. The derivatives which do not appear in the equations will not be a part of the computational graph and will not contribute to any computational costs. Another challenge associated with this inverse problem worth emphasizing is that regions where the concentration field does not vary, or the gradient is zero, gives no information for us to learn and determine the diffusion coefficient. It is noted that automatic differentiation provides us with derivatives calculated with machine precision. We are working with point clouds and just one global function and there is no discretization involved.  This is significant as a lattice like distribution of points is not a requirement and we can work with any distribution of points in the spatio-temporal domain. The popular open-source library Tensorflow \cite{Abadi2016} was used to create and compile the computational graphs for our physics-informed neural network. The parameters for neural networks $\Tilde{c}(x,y,t)$ and $\Tilde{D}(x,y)$ were optimized by minimizing the following mean squared error $(MSE_{Diff})$ function:
    \begin{equation}
        \begin{aligned}
            MSE_{Diff} &= \frac{1}{N} \sum_{i=1}^{N} \left( |\frac{\Tilde{c}(x^{i},y^{i},t^{i})-c(x^{i},y^{i},t^{i})}{\sigma_c}|^{2}\right)\\
           &+ \frac{1}{N^e} \sum_{i=1}^{N^e} \left( |\frac{{c}^{pi}(x^{i},y^{i},t^{i})-\Tilde{c}(x^{i},y^{i},t^{i})}{\sigma_c}|^{2}\right).
        \end{aligned} \label{MSE_diff}
    \end{equation}
 Here $\sigma_c$ is the standard deviation of the concentration, $\{x^{i},y^{i},t^{i}\}_{i=1}^{N}$ denote the points where we have information on $c(x,y,t)$ and the number of such points is $N$, $\{x^{i},y^{i},t^{i}\}_{i=1}^{N^e}$ denote the points where the equation is evaluated on and the number of such points is $N^e$. The second term in eq. \eqref{MSE_diff} is the consistency loss, which ensures that there is consistency in the regressed concentration and the physics informed concentration \cite{Thakur2023}. Both neural networks  $\Tilde{c}(x,y,t)$ and $\Tilde{D}(x,y)$ were densely connected with 8 hidden layers each. The network (Fig. \ref{fig:C-Dnetworks}) was trained with a decaying schedule for the learning rate using the Adam optimizer \cite{Kingma2015}.
 
\section{Results and discussion} \label{Results}

To evaluate the accuracy of our method, the predicted values for concentrations and diffusion coefficients were compared with both data generated from numerical simulations (Section \ref{Numerics}) and  experimentation (Section \ref{Exper}).

\subsection{Numerical Simulations Data}  \label{Numerics}
We considered four different spatially varying diffusion coefficients (referred to as $D_{true}$) as our test cases and numerically solved the diffusion equation with a spatially varying diffusion coefficient (Eq. \eqref{DiffusionEqn}) to get the concentration field of the passive scalar (referred to as $c_{true}$). The PINN framework described in section \ref{Methodology} was then used to determine the spatially varying diffusion coefficient that best described the concentration field distribution. Numerical simulations were carried out using the open-source CFD toolbox OpenFOAM \cite{Weller198} with a 500$\times$500 mesh $(x,y \in (0,1))$ and 100 time steps $(t \in (0,1))$. The mesh point generated from the simulation were used as training points for the PINNs. For all four cases, 5000 points were randomly sampled across time and space from inside the domain for each iteration using the data iterator provided by Tensorflow \cite{Abadi2016}.

For all cases, we considered a square domain and defined the side of the square to be the length scale L and the total time to be the time scale T. The concentration field was scaled using the maximum value of the concentration, hence the dimensionless field $c \in [0,1]$. We defined $x=\frac{x'}{L}$, $y=\frac{y'}{L}$ and $t=\frac{t'}{T}$ where $x'$, $y'$ and $t'$ are the dimensional x-coordinate, y-coordinate and time, respectively while $x$, $y$ and $t$ are the dimensionless x-coordinate, y-coordinate and time, respectively. The diffusion coefficient distribution was scaled by $\frac{T}{L^2}$ to get a dimensionless diffusion coefficient distribution $D(x,y)$. The parameters of the networks were updated using the gradient of the loss with respect to the network parameters to get the optimum set of parameters for the mapping $(x, y, t) \longmapsto (\Tilde{c})$ and $(x, y) \longmapsto (\Tilde{D}) $. For the given concentration distribution, the networks used regression to learn the concentration distribution $\Tilde{c}(x,y,t)$ and minimized the residual defined in equation \eqref{residual_eqn} to determine the diffusion coefficient distribution $\Tilde{D}(x,y)$ which best described the concentration data.

\begin{table}[ht]\label{Result_Summary}
  \caption{Numerically simulated data for the distribution of diffusion coefficient, ${D_{true}}(x,y)$, and the relative error of the predicted to true concentration field and diffusion coefficient, respectively, for each test case.}
  \begin{center}
    \begin{tabular}{|c|c|c|c|}
      \hline
      \bf Case& \bf $D_{true}(x,y)$& \bf $\mathcal{L}(c_{true},\Tilde{c}(x,y,t))$& \bf $\mathcal{L}(D_{true},\Tilde{D}(x,y))$\\
      \hline
      I & $0.05+0.1[x(1-x)+y(1-y)]$ & $1.8\times10^{-4}$& $4.8\times10^{-2}$\\
      \hline
      II & $0.25 + 0.1[sin(2\pi x)+sin(2\pi y)]$& $2.6\times10^{-4}$& $4.6\times10^{-2}$\\
      \hline
      III & $0.25 + 0.1[sin(4\pi x)+sin(4\pi y)]$& $6.78\times10^{-4}$& $6.31\times10^{-2}$\\
      \hline
      IV & $0.3 + 0.15[tanh(-10+20x)] +$ & & \\
      &$0.1 [tanh(-7+20y)]$& $4.81\times10^{-4}$& $4.58\times10^{-2}$\\
      \hline
    \end{tabular}
  \end{center}
  \label{ta:D_summary}
\end{table}

\begin{table}[ht]\label{Result_Summary}
  \caption{The effect of reducing the number of spatial points on the diffusion coefficient for case 1.}
  \begin{center}
    \begin{tabular}{|c|c|c|c|c|c|}
      \hline
      \bf Samples & \bf $250000$& \bf $25000$& \bf $2500$& \bf $250$& \bf $25$\\
      \hline
      Error & $4.89\times10^{-2}$ & $5.9\times10^{-2}$& $6.0\times10^{-2}$& $6.45\times10^{-2}$& $3.9\times10^{-1}$\\
      \hline
      
    \end{tabular}
  \end{center}
  \label{ta:sparse}
\end{table}

\begin{figure*}
    \centering
    \includegraphics[width=1.2\textwidth]{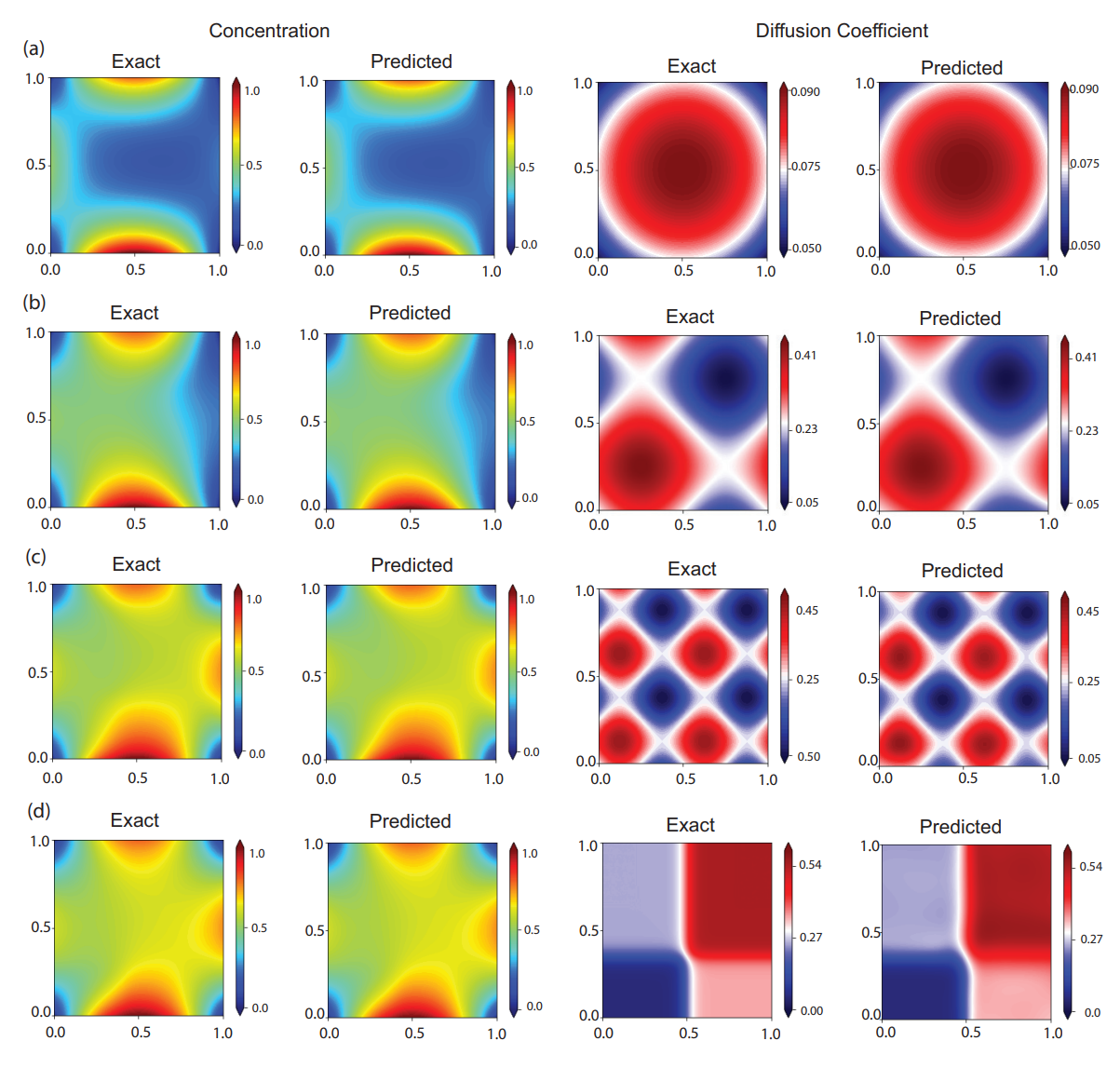}
    \caption{Results for the numerically simulated data. The first column shows the exact concentration distribution, the second column the predicted concentration distribution, the third column shows the exact distribution of the diffusion coefficient and the fourth column shows the predicted distribution of the diffusion coefficient for (a) Case I (b) Case II (c) Case III and (d) Case IV.}
    \label{fig:result_v1}
\end{figure*}

\begin{figure}
    \centering
    \includegraphics[width=0.8\textwidth]{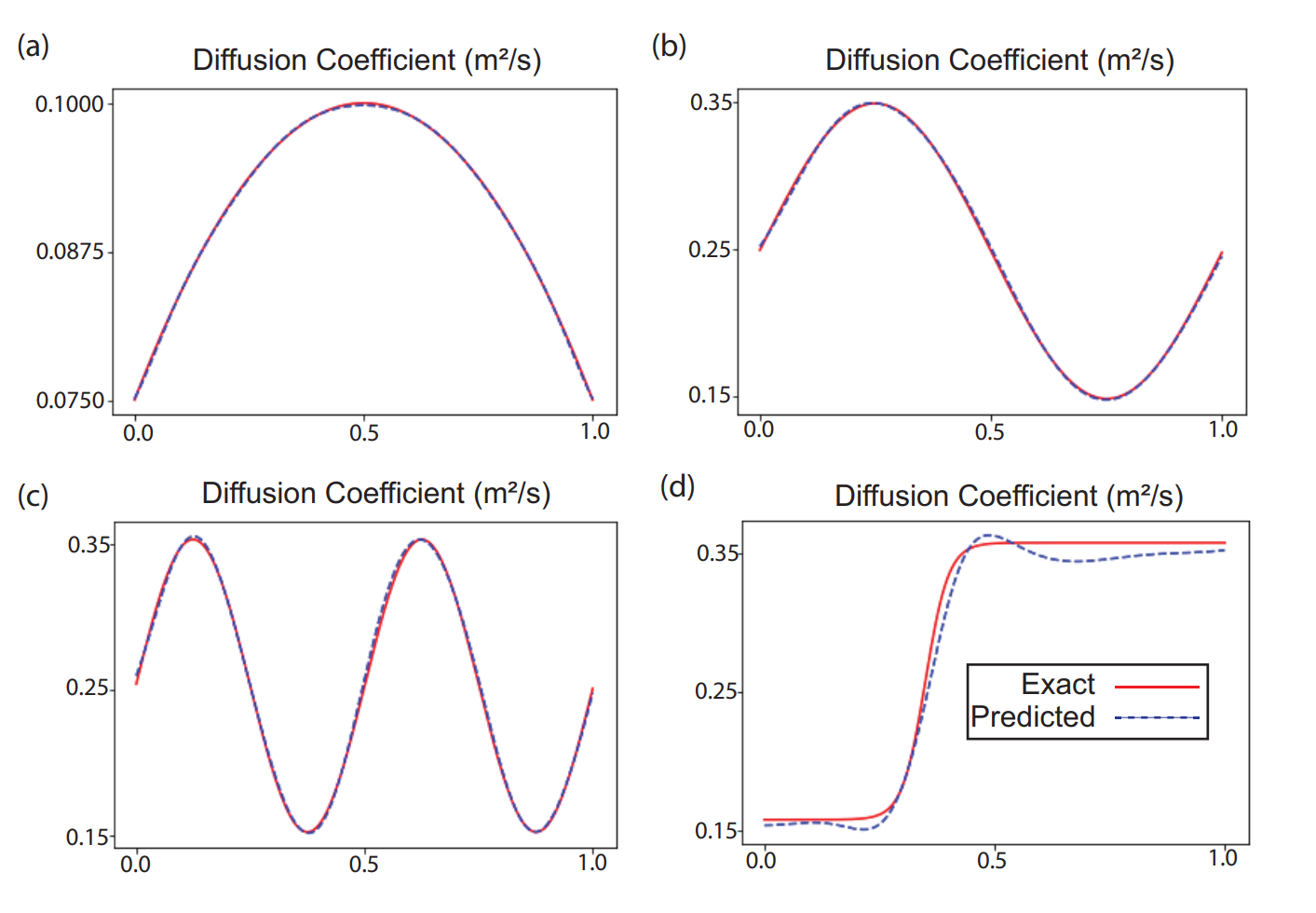}
    \caption{Comparison of the correct (exact) diffusion coefficient and the diffusion coefficient predicted by the neural network along the horizontal cross-section at the midpoint  for (a) Case I (b) Case II (c) Case III and (d) Case IV. The exact value is represented by a solid red line while the prediction of the network is shown by a dotted blue line. }
    \label{fig:result_v2}
\end{figure}

In case I, we simulated the concentration of a passive scalar corresponding to the spatially varying diffusion coefficient $D_{true}(x,y) := 0.05+0.1[x(1-x)+y(1-y)]$. This resulted in a smooth function akin to a two-dimensional parabola. To compare the results predicted by the neural networks to the true value and simulation results, we defined the relative error to be

\begin{equation}
    \mathcal{L}(a,a_{pred}) = \sqrt{\frac{\overline{(a-a_{pred})^2}}{\overline{(a-\Bar{a})^2}}},
\end{equation}
where the bar denotes the mean value. This definition for error is used so that multiplication or addition of a constant does not change the error. In case II, we considered a more complicated two dimensional sinusoidal distribution $D_{true}(x,y) := 0.25 + 0.1[\sin(2\pi x)+\sin(2\pi y)]$ and for case III, we increased the frequency of the sinusoidal function in both directions $D_{true}(x,y) := 0.25 + 0.1[\sin(4\pi x)+\sin(4\pi y)]$. In case IV, we used tangent hyperbolic functions to split the domain into four parts with different diffusion coefficients and continuous boundaries. The diffusion coefficient was defined by $D_{true}(x,y):= 0.3 + 0.15[\tanh(20(-0.5+x))]+ 0.1 [\tanh(20(-0.35+y))]$ for case IV. Such distributions can occur in biological tissues where the underlying tissue microstructure can spatially vary \cite{ning2017}. To choose the penalization parameter, we use the backward Euler discretization and the standard deviation of the concentration field. The concentration network is eight layers deep and each layer has 128 neurons. The same hyperparameters were used for all cases. {Figure \ref{fig:result_v1}} shows a snapshot of the results from the network at $t=0.5T$ for all the four cases. The first column shows the exact value of the concentration (${c}(x,y,0.5T)$) and the second column shows the value predicted by the neural network ($\Tilde{c}(x,y,0.5T)$). The third column in {figure \ref{fig:result_v1}} shows the true value of the diffusion coefficient (${D}(x,y)$) while the fourth column shows the value predicted by the neural network ($\Tilde{D}(x,y)$).  The relative error between the learned concentration and diffusion coefficient distributions and the true values from the simulation are summarized in Table \ref{ta:D_summary}. Our method learned the diffusion coefficient reliably in all cases, with the relative error being below $6.31\times10^{-2}$ in all cases. 
{Figure \ref{fig:result_v2}} compares the exact diffusion coefficient and the diffusion coefficient predicted by the neural network in all the four cases along the horizontal cross-section at the midpoint. To test the robustness of our method to sparse data, we reduce the number of spatial points as observations to the concentration field systematically for case 1. We look at 250000, 25000, 2500, 250 and 25 spatial data points and the results are shared in table \ref{ta:sparse}. The error in the diffusion coefficient only marginally increases till the number of points are reduced to 250. The solver finally breaks down when only 25 spatial points or “pixels” are considered. This demonstrates that our framework is not sensitive to resolution of data. With the method performing well with noiseless data, we next considered images from experiments. 

\subsection{Experimental Data}\label{Exper}
To verify our algorithm on experimental data, we obtained time lapse images of Rhodamine 6G dye diffusing through (1) two liquids (Section \ref{Diff_liq}), (2) agarose, and (3) gelatin hydrogels (Section \ref{Diff_gels}). The intensity of the images was assumed to be proportional to the concentration of the dye for our calculations. Working with experimental data is significantly more challenging compared to numerical data as various sources of noise contribute to perturbations in the observations. These perturbations can significantly change the estimated diffusion coefficient \cite{Yang2008}. Neural networks can be effective tools for processing the images and mitigating noise effects. A neural network can be trained on a regression task over the concentration data and automatic differentiation can be used to visualize the temporal and spatial derivatives. This helps in identification and removal of aberrations and helps in understanding the data better. As part of processing the images, the space and the time domains were non-dimensionalized using an appropriate length and time scale.

\subsubsection{Diffusion through Liquid}\label{Diff_liq}
\begin{figure*}
    \centering
    \includegraphics[width=1.25\textwidth]{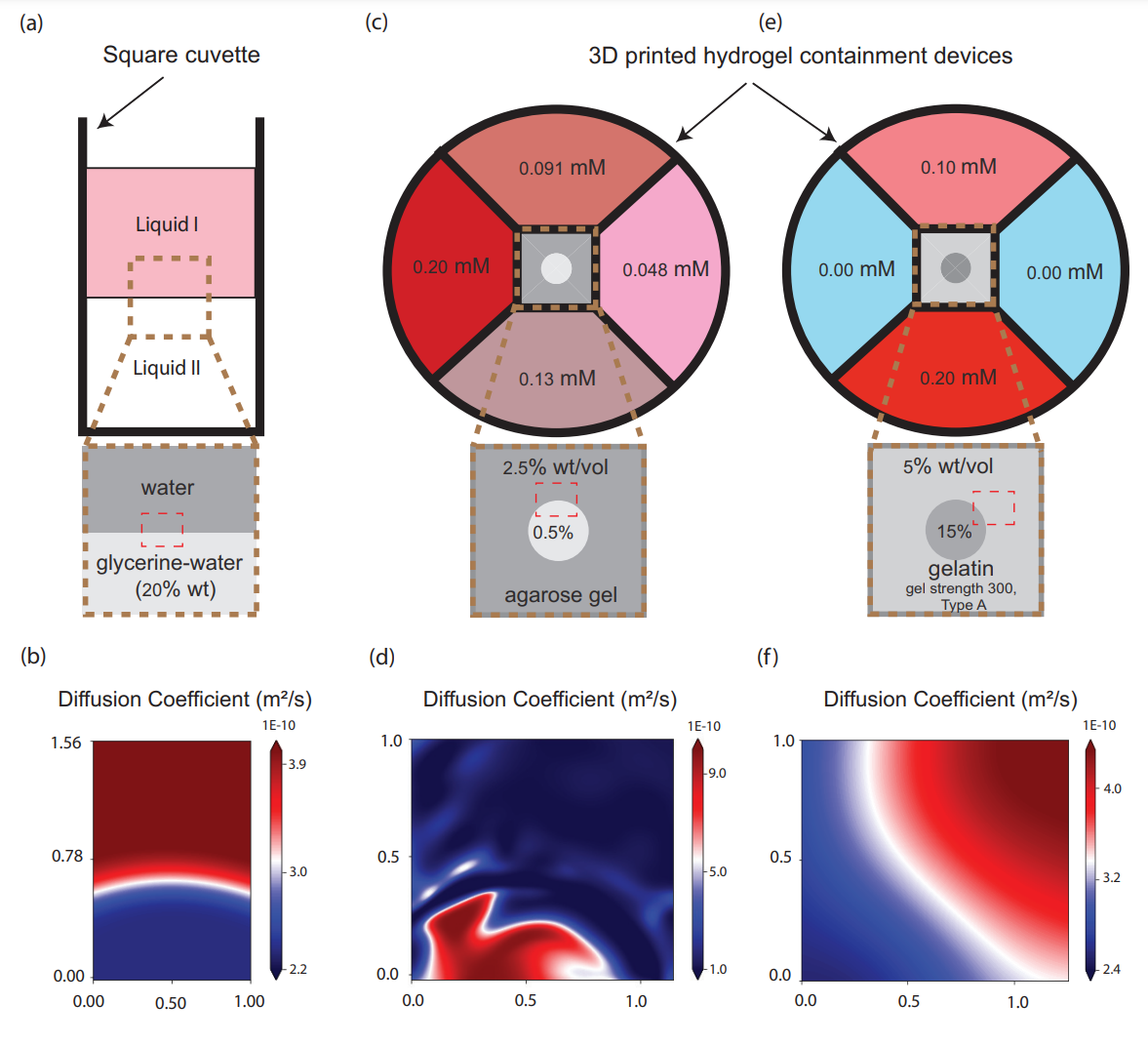}
    \caption{Experimental setup and results for diffusion of Rhodamine 6G dye. In each setup, the approximate areas used for diffusion analyses are denoted with red boxes. (a) Different liquid solutions in a square cuvette (DI water and  glycerine-water mixture). (b) The neural network prediction for the diffusion coefficient for water and water-Glycerine mixture. (c) Initial source concentrations of Rhodamine 6G dye in PBS to diffuse into a 2.5\% and 0.5\% agarose gel chamber. (d) The neural network prediction for the diffusion coefficient for diffusion through the agarose gel. (e) Initial source concentrations of Rhodamine 6G dye in PBS to diffuse into a 5\% and 15\% gelatin chamber. (f) The neural network prediction for the diffusion coefficient for diffusion through gelatin the gel. The x and y coordinates were made dimensionless using appropriate length scales. }
    \label{fig:exp_diff}
\end{figure*}

The setup, shown in  Fig. \ref{fig:exp_diff}(a), was designed to investigate the diffusion coefficients of different liquid solutions. A square cuvette [Polymethyl methacrylate (PMMA) cuvette] with dimensions of 1.25$\times$1.25$\times$4.5 cm was filled with two layers of solutions: DI water (liquid {1}) as an upper layer and a glycerine-water mixture (20$\%$ wt, liquid {2}) in a lower layer. The water solution on top contained 100 $\mu$M  Rhodamine 6g dye (Sigma-Aldrich, Rhodamine 6G, dye content 99\%,  $C_{28}H_{31}N_{2}O_{3}Cl$, molecular weight 479.01 g/mol, SKU 252433) and over time, diffusion occurred between the two layers. A camera (Imperx, CLM-B6640M-TF000) was used to capture the diffusion every minute with a resolution of 6640$\times$4400 pixels. The images were analyzed using a custom Matlab algorithm to measure the dye intensity in each frame. The setup was placed inside a photobox to prevent ambient light interference and the temperature was maintained at 23$^{\circ}$ Celsius.

Different values have been reported for the diffusion coefficient of Rhodamine 6G \cite{gendron2008,majer2014}. Here we used the following viscosity values of $\mu_{1}=9.35 \times 10^{-4}$ Ns/m$^2$ and $\mu_{2}=1.60 \times 10^{-3}$ Ns/m$^2$ for the respective solutions and calculated the diffusion coefficient at room temperature (using Stokes-Einstein equation) as $D_{1}= 4.05 \times 10^{-10}$ m$^2$/s and $D_{{2}}= 2.36 \times 10^{-10}$ m$^2$/s \cite{gendron2008,majer2014,culbertson2002diffusion}.   
{For this case, the neural network prediction for the spatially varying diffusion coefficient is shown in Fig. \ref{fig:exp_diff} (b). We have used a cuvette to look at the diffusion between liquids. The images were then captured using a camera and then processed through a solver. We believe that the curvature has been caused as the interface between the two fluids isn’t perfectly flat and due to the distortion effects of approximating the three-dimensional cuvette using two-dimensional images. Two distinct regions are visible in the plot and the diffusion coefficient from the neural network has the value $ 4.0 \times 10^{-10}$ m$^2$/s and $ 2.4 \times 10^{-10}$ m$^2$/s for the water and glycerine-water regions, respectively, which are in good agreement with the calculated ground truth values and with previously-reported literature using numerous techniques \cite{gendron2008,majer2014, culbertson2002diffusion}. }

\subsubsection{Diffusion through Gels}\label{Diff_gels}

Rectangular prisms of either gelatin or agarose hydrogels were fabricated such that varied concentrations of Rhodamine 6G dye in phosphate buffered saline (PBS) would diffuse inward from each side, mimicking the computational setup. A custom computer-aided design (CAD) part was drawn in Autodesk Inventor and fabricated using Flexible Resin V1 (RS-F2-FLGR-02) on the Formlabs Form 3 SLA 3D printer. The part was made to fit the diameter of the CELLSTAR 35x10mm cell culture petri dish (Cat. No. 627-160), segmenting the dish into four equal reservoirs with an open container in the center where the hydrogel was situated. In both experiments, the hydrogels were 4 mm\textsuperscript{2} by 5 mm tall rectangular prisms, dictated by the container size of the 3D printed part. The windows on the sides of the gel that contacted the PBS/dye baths were 3 mm\textsuperscript{2} squares. Directly before the kinetic experiment began, different volumes of a 1 mM Rhodamine 6G stock solution were added into the different PBS reservoirs, resulting in the source dye concentrations shown in Fig. \ref{fig:exp_diff}(c,e). Brightfield and red fluorescence images (BioTek Cytation 5 RFP, 531, 593 filter cube and LED cube set) were captured on the BioTek Cytation 5 using the Gen5 software with the BioTek 4x Plan Fluorite phase objective (BioTek, 1320515).

 In the agarose experiments, 2.5\% wt/vol agarose (Fisher BioReagents, BP160500) in PBS was flooded into the hydrogel container while a plastic cylinder, 1.5 mm in diameter, was positioned in the center of the container as a place-holder. Once the 2.5\% agarose gelled completely, the cylinder was removed and 0.5\% wt/vol agarose was transferred into the resulting hole and allowed to solidify (Fig. \ref{fig:exp_diff}(c)). Brightfield and red fluorescent images were obtained every minute. 
 
 For gelatin experiments, gelatin from porcine skin, gel strength 300, Type A (Sigma Life Sciences, G2500-500G) was used at 5\% and 15\% wt/vol. Because of the nature of gelatin, the same fabrication technique could not be used as above. Instead, 15\% gelatin in PBS was drawn up into the luer slip tip of a 1 mL syringe (approximately 2.2 mm in diameter) (Henke Sass Wolf, 4010-200V0) and allowed to gel. The gelatin cylinder was then placed into the 3D printed part and 5\% gelatin was flooded around the rest of the container and allowed to gel (Fig. \ref{fig:exp_diff}(e)). Due to the slower diffusion through gelatin which required longer kinetic runs, brightfield and red fluorescent images were obtained every 2.5 minutes.

In both hydrogel cases, the neural network prediction was able to identify two distinct regions of spatially-varying diffusion coefficients corresponding to the various formulations of the hydrogels. Within the agarose gel, values of approximately $ 2.0 \times 10^{-10}$ m$^2$/s and $ 9.0 \times 10^{-10}$ m$^2$/s were obtained for the 2.5\% and 0.5\% wt/vol agarose regions, respectively (Fig. \ref{fig:exp_diff}(d)). Similarly, the values for the diffusion coefficients in the two gelatin regions were roughly $2.4 \times 10^{-10}$ m$^2$/s and $3.7 \times 10^{-10}$ m$^2$/s for the 15\% and 5\% wt/vol gelatin regions, respectively (Fig. \ref{fig:exp_diff}(f)). Because our experimental setup introduces a varied concentration gradient of Rhodamine 6G at each interface of the hydrogel, the obtained diffusion coefficients will not be identical to literature values that use saturated gels, such as the common Fluorescence Recovery After Photobleaching (FRAP) technique. However, we observed the expected trend where both the 2.5\% and 15\% gels had lower diffusion coefficient values than their less-dense counterparts. Additionally, the range of values obtained is consistent with those found in the literature for similar setups. Golmohamadi et al. found the diffusion coefficient of Rhodamine 6G to be approximately $ 2.5-3 \times 10^{-10}$ m$^2$/s at neutral pH in 1.5\% agarose gels \cite{golmohamadi2012diffusion}. Samprovalaki et al. similarly found the diffusion coefficient of Rhodamine 6G to be around $ 2.89-4.88 \times 10^{-10}$ m$^2$/s at 30 degree Celsius in 1.5\% agar gel depending on the concentration of Rhodamine 6G used \cite{samprovalaki2012investigation}. Within gelatin, Stucchi et al. observed a diffusion coefficient for Rhodamine 6G of $ 4 \times 10^{-11}$ m$^2$/s for 5\% gelatin hydrogels. However, these hydrogels were crosslinked using diethyl squarate to increase the gel density and other mechanical properties, which would result in a slower rate of diffusion than the unmodified gelatin used here \cite{stucchi2021squarate}. Additionally, the diffusion maps in Fig. \ref{fig:exp_diff}(d,f) were able to capture the varied boundary qualities between the agarose and gelatin samples. The agarose results indicate a more jagged boundary between the gel concentrations, likely resulted from the removal of the plastic cylinder after the 2.5\% agarose had gelled around it. This is not apparent with the gelatin results (Fig. \ref{fig:exp_diff}(d)) after the fabrication method was altered such that the cylinder of 15\% gelatin was extruded separately, placed within the custom-printed holder, and flooded with 5\% gelatin. The jagged lines that we observe are due to the nature of the experiment, where distinct gels were artificially connected to mimic spatially varying diffusion. Hence, the boundaries are not smooth and lead to sudden variation in the diffusion coefficient. We don’t expect to see such variations where anomalous diffusion is naturally present and the underlying structure is inherently heterogenous. Taken together, these results provide strong evidence that the algorithm presented is accurate for several types of experimental data that may utilize distinct materials and hardware to capture rates of diffusion. 

\section{Conclusion}\label{conclusion}
 The diffusion coefficient often varies spatially in biological tissues, porous media, and soil. In this work, we used a physics informed neural network (PINN) to solve the inverse problem of discovering the spatially varying diffusion coefficient from spatio-temporal information on the diffused passive scalar. The results of the framework were first validated and the accuracy of the method was tested using noiseless, numerically-generated data. Inverse problems are notoriously difficult to solve as they are often ill-posed and small perturbations can significantly change the estimated diffusion coefficients. During experiments, multiple sources of noise create such perturbations in the recorded data. To demonstrate the robustness of the framework, we then considered experimental data as Rhodamine 6G dye diffused through different liquids and hydrogels.  Specifically, for diffusion through liquids, diffusion of the dye was captured from a water layer into a layer of a water-glycerine mixture. For diffusion through gels, Rhodamine 6G dye was diffused through either 2.5\% and 0.5\% wt/vol heterogeneous agarose gels, or 5\% and 15\% wt/vol heterogeneous gelatin hydrogels. It was shown that our framework captured the spatial distribution of the diffusion coefficient and the interface of distinct regions well in each test case. Using this methodology can thus allow the prediction of the spatiotemporal distribution and variation inside biological samples, where understanding the heterogeneity is crucial. This has implications to aid numerous biological fields and could allow for more accurate calculation of the diffusion of growth factors or other therapeutics from implanted devices, or contribute to a  diagnosis of aberrations such as carcinogenic tumours in tissues using data acquired through medical imaging, as two examples. 
\section{Author Contributions}
ST performed computational research and analyzed data. EE, SL, and LS designed and performed experimental research. ST, EE, SL, LS, and AMA  wrote the paper. ST and AMA designed research.

\section{Declaration of Interests}
The authors declare no competing interests.

\section{Acknowledgements}
The authors would like to thank \'Angel Enr\'iquez for operating the 3D printer to fabricate the custom hydrogel divider and Andres Barrio-Zhang for helping with the experimental setup. This research was supported in part through the National Science Foundation (CBET-1700961 and CBET 1705371) to AMA and the CTSI TL1 Predoctoral Training Fellowship (UL1TR002529) and the National Cancer Institute Predoctoral to Postdoctoral Fellowship Transition Award (F99CA264734) to SL. 

\bibliographystyle{elsarticle-num}
\bibliography{main}

\begin{thebibliography}{10}
\expandafter\ifx\csname url\endcsname\relax
  \def\url#1{\texttt{#1}}\fi
\expandafter\ifx\csname urlprefix\endcsname\relax\def\urlprefix{URL }\fi
\expandafter\ifx\csname href\endcsname\relax
  \def\href#1#2{#2} \def\path#1{#1}\fi

\bibitem{bruna2015}
T.~O. Optimality, {Diffusion in spatially varying porous media} 25~(1) (2015) 76--101.

\bibitem{ning2017}
L.~Ning, E.~{\"{O}}zarslan, C.~F. Westin, Y.~Rathi, \href{http://dx.doi.org/10.1016/j.neuroimage.2016.09.057}{{Precise Inference and Characterization of Structural Organization (PICASO) of tissue from molecular diffusion}}, NeuroImage 146~(September 2016) (2017) 452--473.
\newblock \href {https://doi.org/10.1016/j.neuroimage.2016.09.057} {\path{doi:10.1016/j.neuroimage.2016.09.057}}.
\newline\urlprefix\url{http://dx.doi.org/10.1016/j.neuroimage.2016.09.057}

\bibitem{Kabanikhin2008}
S.~I. Kabanikhin, {Definitions and examples of inverse and ill-posed problems}, Journal of inverse and Ill-posed problems 16~(4) (2008) 317--357.

\bibitem{Yang2008}
L.~Yang, J.~N. Yu, Z.~C. Deng, {An inverse problem of identifying the coefficient of parabolic equation}, Applied Mathematical Modelling 32~(10) (2008) 1984--1995.
\newblock \href {https://doi.org/10.1016/j.apm.2007.06.025} {\path{doi:10.1016/j.apm.2007.06.025}}.

\bibitem{Kern2016}
M.~Kern, {Numerical Methods for Inverse Problems}, John Wiley {\&} Sons, 2016.

\bibitem{Isakov2006}
V.~Isakov, {Inverse problems for partial differential equations}, Vol. 127, Springer, 2006.

\bibitem{Shifdar2006}
A.~Shidfar, R.~Pourgholi, M.~Ebrahimi, {A numerical method for solving of a nonlinear inverse diffusion problem}, Computers {\&} Mathematics with Applications 52~(6-7) (2006) 1021--1030.

\bibitem{Mazraeh2017}
H.~D. Mazraeh, R.~Pourgholi, T.~Houlari, {Combining genetic algorithm and sinc-galerkin method for solving an inverse diffusion problem} 7~(1) (2017) 33--50.

\bibitem{Fudym2008}
O.~Fudym, H.~R. Orlande, M.~Bamford, J.~C. Batsale, {Bayesian approach for thermal diffusivity mapping from infrared images with spatially random heat pulse heating}, Journal of Physics: Conference Series 135~(Mcmc) (2008).
\newblock \href {https://doi.org/10.1088/1742-6596/135/1/012042} {\path{doi:10.1088/1742-6596/135/1/012042}}.

\bibitem{boodaghi2021}
M.~Boodaghi, S.~Libring, L.~Solorio, A.~M. Ardekani, \href{https://doi.org/10.1016/j.jconrel.2021.10.002}{{A Bayesian approach to estimate the diffusion coefficient of Rhodamine 6G in breast cancer spheroids}}, Journal of Controlled Release 340~(May) (2021) 60--71.
\newblock \href {https://doi.org/10.1016/j.jconrel.2021.10.002} {\path{doi:10.1016/j.jconrel.2021.10.002}}.
\newline\urlprefix\url{https://doi.org/10.1016/j.jconrel.2021.10.002}

\bibitem{Lusch2018}
B.~Lusch, J.~N. Kutz, S.~L. Brunton, {Deep learning for universal linear embeddings of nonlinear dynamics}, Nature Communications 9~(1) (2018).
\newblock \href {https://doi.org/10.1038/s41467-018-07210-0} {\path{doi:10.1038/s41467-018-07210-0}}.

\bibitem{Brunton2020}
S.~L. Brunton, B.~R. Noack, P.~Koumoutsakos, {Machine Learning for Fluid Mechanics}, Annual Review of Fluid Mechanics 52~(1) (2020) 477--508.
\newblock \href {https://doi.org/10.1146/annurev-fluid-010719-060214} {\path{doi:10.1146/annurev-fluid-010719-060214}}.

\bibitem{Sanchez2018}
B.~Sanchez-Lengeling, A.~Aspuru-Guzik, {Inverse molecular design using machine learning:Generative models for matter engineering}, Science 361~(6400) (2018) 360--365.
\newblock \href {https://doi.org/10.1126/science.aat2663} {\path{doi:10.1126/science.aat2663}}.

\bibitem{Parish2016}
E.~J. Parish, K.~Duraisamy, \href{http://dx.doi.org/10.1016/j.jcp.2015.11.012}{{A paradigm for data-driven predictive modeling using field inversion and machine learning}}, Journal of Computational Physics 305 (2016) 758--774.
\newblock \href {https://doi.org/10.1016/j.jcp.2015.11.012} {\path{doi:10.1016/j.jcp.2015.11.012}}.
\newline\urlprefix\url{http://dx.doi.org/10.1016/j.jcp.2015.11.012}

\bibitem{Ling2016}
J.~Ling, A.~Kurzawski, J.~Templeton, {Reynolds averaged turbulence modelling using deep neural networks with embedded invariance}, Journal of Fluid Mechanics 807 (2016) 155--166.
\newblock \href {https://doi.org/10.1017/jfm.2016.615} {\path{doi:10.1017/jfm.2016.615}}.

\bibitem{Writh2021}
D.~Wirth, A.~McCall, K.~Hristova, \href{https://doi.org/10.1016/j.bpj.2021.04.030}{{Neural network strategies for plasma membrane selection in fluorescence microscopy images}}, Biophysical Journal 120~(12) (2021) 2374--2385.
\newblock \href {https://doi.org/10.1016/j.bpj.2021.04.030} {\path{doi:10.1016/j.bpj.2021.04.030}}.
\newline\urlprefix\url{https://doi.org/10.1016/j.bpj.2021.04.030}

\bibitem{AutoTracer2021}
M.~Schneider, A.~Al-Shaer, N.~R. Forde, \href{https://doi.org/10.1016/j.bpj.2021.05.011}{{AutoSmarTrace: Automated chain tracing and flexibility analysis of biological filaments}}, Biophysical Journal 120~(13) (2021) 2599--2608.
\newblock \href {https://doi.org/10.1016/j.bpj.2021.05.011} {\path{doi:10.1016/j.bpj.2021.05.011}}.
\newline\urlprefix\url{https://doi.org/10.1016/j.bpj.2021.05.011}

\bibitem{Raissi2017}
M.~Raissi, P.~Perdikaris, G.~E. Karniadakis, \href{http://arxiv.org/abs/1711.10561}{{Physics Informed Deep Learning (Part I): Data-driven Solutions of Nonlinear Partial Differential Equations}}~(Part I) (2017) 1--22.
\newline\urlprefix\url{http://arxiv.org/abs/1711.10561}

\bibitem{Raissi2017a}
M.~Raissi, P.~Perdikaris, G.~E. Karniadakis, \href{http://arxiv.org/abs/1711.10566}{{Physics Informed Deep Learning (Part II): Data-driven Discovery of Nonlinear Partial Differential Equations}}~(Part II) (2017) 1--19.
\newline\urlprefix\url{http://arxiv.org/abs/1711.10566}

\bibitem{Raissi2019}
M.~Raissi, P.~Perdikaris, G.~E. Karniadakis, \href{https://doi.org/10.1016/j.jcp.2018.10.045}{{Physics-informed neural networks: A deep learning framework for solving forward and inverse problems involving nonlinear partial differential equations}}, Journal of Computational Physics 378 (2019) 686--707.
\newblock \href {https://doi.org/10.1016/j.jcp.2018.10.045} {\path{doi:10.1016/j.jcp.2018.10.045}}.
\newline\urlprefix\url{https://doi.org/10.1016/j.jcp.2018.10.045}

\bibitem{Lu2021}
L.~Lu, X.~Meng, Z.~Mao, G.~E. Karniadakis, {DeepXDE: A deep learning library for solving differential equations}, SIAM Review 63~(1) (2021) 208--228.
\newblock \href {https://doi.org/10.1137/19M1274067} {\path{doi:10.1137/19M1274067}}.

\bibitem{Goswami2019}
S.~Goswami, C.~Anitescu, S.~Chakraborty, T.~Rabczuk, \href{https://doi.org/10.1016/j.tafmec.2019.102447}{{Transfer learning enhanced physics informed neural network for phase-field modeling of fracture}}, Theoretical and Applied Fracture Mechanics 106~(July 2019) (2019) 102447.
\newblock \href {https://doi.org/10.1016/j.tafmec.2019.102447} {\path{doi:10.1016/j.tafmec.2019.102447}}.
\newline\urlprefix\url{https://doi.org/10.1016/j.tafmec.2019.102447}

\bibitem{Tipireddy2019}
R.~Tipireddy, P.~Perdikaris, P.~Stinis, A.~Tartakovsky, \href{http://arxiv.org/abs/1904.04058}{{A comparative study of physics-informed neural network models for learning unknown dynamics and constitutive relations}} (2019).
\newline\urlprefix\url{http://arxiv.org/abs/1904.04058}

\bibitem{niaki2021}
S.~Amini~Niaki, E.~Haghighat, T.~Campbell, A.~Poursartip, R.~Vaziri, \href{https://doi.org/10.1016/j.cma.2021.113959}{{Physics-informed neural network for modelling the thermochemical curing process of composite-tool systems during manufacture}}, Computer Methods in Applied Mechanics and Engineering 384 (2021) 113959.
\newblock \href {https://doi.org/10.1016/j.cma.2021.113959} {\path{doi:10.1016/j.cma.2021.113959}}.
\newline\urlprefix\url{https://doi.org/10.1016/j.cma.2021.113959}

\bibitem{kissas2020}
G.~Kissas, Y.~Yang, E.~Hwuang, W.~R. Witschey, J.~A. Detre, P.~Perdikaris, \href{https://doi.org/10.1016/j.cma.2019.112623}{{Machine learning in cardiovascular flows modeling: Predicting arterial blood pressure from non-invasive 4D flow MRI data using physics-informed neural networks}}, Computer Methods in Applied Mechanics and Engineering 358 (2020) 112623.
\newblock \href {https://doi.org/10.1016/j.cma.2019.112623} {\path{doi:10.1016/j.cma.2019.112623}}.
\newline\urlprefix\url{https://doi.org/10.1016/j.cma.2019.112623}

\bibitem{liu2020}
M.~Liu, L.~Liang, W.~Sun, \href{https://doi.org/10.1016/j.cma.2020.113402}{{A generic physics-informed neural network-based constitutive model for soft biological tissues}}, Computer Methods in Applied Mechanics and Engineering 372 (2020) 113402.
\newblock \href {https://doi.org/10.1016/j.cma.2020.113402} {\path{doi:10.1016/j.cma.2020.113402}}.
\newline\urlprefix\url{https://doi.org/10.1016/j.cma.2020.113402}

\bibitem{Raissi2020}
M.~Raissi, A.~Yazdani, G.~E. Karniadakis, \href{https://science.sciencemag.org/content/367/6481/1026/tab-pdf}{{Hidden fluid mechanics: Learning velocity and pressure fields from flow visualizations}}~(C) (2020) 1--5.
\newblock \href {https://doi.org/10.1126/science.aaw4741} {\path{doi:10.1126/science.aaw4741}}.
\newline\urlprefix\url{https://science.sciencemag.org/content/367/6481/1026/tab-pdf}

\bibitem{Raissi2019a}
M.~Raissi, H.~Babaee, P.~Givi, {Deep learning of turbulent scalar mixing}, Physical Review Fluids 4~(12) (2019) 1--19.
\newblock \href {https://doi.org/10.1103/PhysRevFluids.4.124501} {\path{doi:10.1103/PhysRevFluids.4.124501}}.

\bibitem{Abadi2016}
M.~Abadi, A.~Agarwal, P.~Barham, E.~Brevdo, Z.~Chen, C.~Citro, G.~S. Corrado, A.~Davis, J.~Dean, M.~Devin, S.~Ghemawat, I.~Goodfellow, A.~Harp, G.~Irving, M.~Isard, Y.~Jia, R.~Jozefowicz, L.~Kaiser, M.~Kudlur, J.~Levenberg, D.~Mane, R.~Monga, S.~Moore, D.~Murray, C.~Olah, M.~Schuster, J.~Shlens, B.~Steiner, I.~Sutskever, K.~Talwar, P.~Tucker, V.~Vanhoucke, V.~Vasudevan, F.~Viegas, O.~Vinyals, P.~Warden, M.~Wattenberg, M.~Wicke, Y.~Yu, X.~Zheng, \href{http://arxiv.org/abs/1603.04467}{{TensorFlow: Large-Scale Machine Learning on Heterogeneous Distributed Systems}} (2016).
\newline\urlprefix\url{http://arxiv.org/abs/1603.04467}

\bibitem{Thakur2023}
S.~Thakur, M.~Raissi, H.~Mitra, A.~Ardekani, \href{http://arxiv.org/abs/2301.13262}{{Temporal Consistency Loss for Physics-Informed Neural Networks}} (2023) 1--14.
\newline\urlprefix\url{http://arxiv.org/abs/2301.13262}

\bibitem{Kingma2015}
D.~P. Kingma, J.~L. Ba, {Adam: A method for stochastic optimization}, 3rd International Conference on Learning Representations, ICLR 2015 - Conference Track Proceedings (2015) 1--15.

\bibitem{Weller198}
H.~G. Weller, G.~Tabor, H.~Jasak, C.~Fureby, {A tensorial approach to computational continuum mechanics using object-oriented techniques}, Computers in Physics 12~(6) (1998) 620.
\newblock \href {https://doi.org/10.1063/1.168744} {\path{doi:10.1063/1.168744}}.

\bibitem{gendron2008}
P.~O. Gendron, F.~Avaltroni, K.~J. Wilkinson, {Diffusion coefficients of several rhodamine derivatives as determined by pulsed field gradient-nuclear magnetic resonance and fluorescence correlation spectroscopy}, Journal of Fluorescence 18~(6) (2008) 1093--1101.
\newblock \href {https://doi.org/10.1007/s10895-008-0357-7} {\path{doi:10.1007/s10895-008-0357-7}}.

\bibitem{majer2014}
G.~Majer, J.~P. Melchior, {Characterization of the fluorescence correlation spectroscopy (FCS) standard Rhodamine 6G and calibration of its diffusion coefficient in aqueous solutions}, Journal of Chemical Physics 140~(9) (2014).
\newblock \href {https://doi.org/10.1063/1.4867096} {\path{doi:10.1063/1.4867096}}.

\bibitem{culbertson2002diffusion}
C.~T. Culbertson, S.~C. Jacobson, J.~Michael~Ramsey, {Diffusion coefficient measurements in microfluidic devices}, Talanta 56~(2) (2002) 365--373.
\newblock \href {https://doi.org/10.1016/S0039-9140(01)00602-6} {\path{doi:10.1016/S0039-9140(01)00602-6}}.

\bibitem{golmohamadi2012diffusion}
M.~Golmohamadi, T.~A. Davis, K.~J. Wilkinson, {Diffusion and partitioning of cations in an agarose hydrogel}, Journal of Physical Chemistry A 116~(25) (2012) 6505--6510.
\newblock \href {https://doi.org/10.1021/jp212343g} {\path{doi:10.1021/jp212343g}}.

\bibitem{samprovalaki2012investigation}
K.~Samprovalaki, P.~T. Robbins, P.~J. Fryer, \href{http://dx.doi.org/10.1016/j.jfoodeng.2012.03.024}{{Investigation of the diffusion of dyes in agar gels}}, Journal of Food Engineering 111~(4) (2012) 537--545.
\newblock \href {https://doi.org/10.1016/j.jfoodeng.2012.03.024} {\path{doi:10.1016/j.jfoodeng.2012.03.024}}.
\newline\urlprefix\url{http://dx.doi.org/10.1016/j.jfoodeng.2012.03.024}

\bibitem{stucchi2021squarate}
S.~Stucchi, D.~Colombo, R.~Guizzardi, A.~D'Aloia, M.~Collini, M.~Bouzin, B.~Costa, M.~Ceriani, A.~Natalello, P.~Pallavicini, L.~Cipolla, {Squarate Cross-Linked Gelatin Hydrogels as Three-Dimensional Scaffolds for Biomedical Applications}, Langmuir 37~(48) (2021) 14050--14058.
\newblock \href {https://doi.org/10.1021/acs.langmuir.1c02080} {\path{doi:10.1021/acs.langmuir.1c02080}}.

\end{thebibliography}
\end{document}